# Modeling of Electrochemical Oxide Film Growth – Impact of Band-to-Band Tunneling

Ingmar Bösing*[1,3], Jorg Thöming[1,3], Fabio La Mantia[2,3]

[1]Chemical Process Engineering Group (CVT)
[2]Energy storage and conversion systems
[3]University of Bremen
*corresponding author, ingmarboesing@uni-bremen.de
Leobener Strasse 6
28359 Bremen
Germany

Keywords: Passivity, metal oxides, modeling, Point Defect Model, Band-to-band tunneling

*Abstract* The Point Defect Model (PDM) describes the corrosion resistance properties of oxide films based on interfacial reactions and defect transport, which are affected by the electric field inside the oxide film. The PDM assumes a constant electric field strength due to band-to-band tunneling (BTBT) of electrons and the separation of electrons and holes by high electric fields. In this manuscript we present a more complex expansion of the common models to simulate steady state oxide films to test this assumption. The R-PDM was extended by including the transport of electrons and holes and BTBT. It could be shown that BTBT only occurs in very rare cases of narrow band gaps and high electric fields and the impact of electrons and holes does indeed lead to a buffering effect on the electric field, but does not lead to a constant electric field strength. Modeling the transport of electrons and holes on the oxide film allows to specifically estimate their potential impact on the film growth. Especially during modeling of oxide films with narrow band gap and/or electrochemical reactions at the film/solution interface the electrons and holes needs to be included to the model.

## 1. INTRODUCTION

The Point Defect Model (PDM) is developed in 1981 to describe oxide film growth by a set of interfacial reactions and transport processes [1]. Since the early beginnings the PDM has seen several changes and extensions to enable the description of oxide film growth on several metals [2]. Furthermore, it has been shown that the PDM fits experimental data very well, and can be used to extract kinetic parameters from impedance data [3–9].

Despite the extensive use of the model and the possibility of depicting experimental data via the model the PDM is still critically discussed. One of the main assumptions of the PDM – a constant electrical field strength independent of external potential inside the oxide – was often a matter of debate [10–12]. Even though, the referenced literature did expand the PDM to achieve a deeper insight into the behavior of the electrical field under potentiostatic control, the hypothesis of the PDM (buffering of the electric field by band-to-band tunneling [2]), was not included to the models. Thus, the hypothesis is not tested and the potential effect of BTBT on the modelled electric field distribution is not described until now.

The simulated steady state behavior of oxide films, described by taking into account only defects and by excluding electrons and holes, has been extensively discussed in our previous publication [13]. It could be shown that the electric field distribution inside the oxide film depends on the external potential and on the position inside the film. Since, on the one hand, the electric field is affected by the charge carriers (oxygen and metal vacancies in the discussed case), as can be seen by the Poisson equation, but on the other hand does affect the transport and generation of the charge carriers (and thus the steady state film thickness), by its influence on the potential, a deep knowledge of possible effects on the electric field is important for the modeling of oxide films.

As pointed out by Macdonald et al. the electric field can also be affected by electrons and holes [2], thus, their influence on the electric field needs to be discussed. As Macdonald et al. argued, high electric fields can lead to band bending, which enables band-to-band tunneling (BTBT) and thereby to a high amount of additional charge carriers (electrons in the conduction band and holes in the valence band). By the electric field inside the oxide film both charge carriers can be separated. This in turn might buffer the electric field at values between 2-5 MV/cm, an effect that could lead to a potential independent electric field inside the oxide [2].

There are already some attempts to implement the impact of electrons and holes to the modeling of oxide films. In 2010 Bataillon et al. presented a comprehensive model for oxide film growth on iron, called The Diffusion Poisson Coupled Model, including the transport of electrons [14]. In contrast to the present study, they did not include the transport and generation of holes and the possibility of band-to-band tunneling. Furthermore, they used a different approach for the boundary conditions of their model, based on the differential capacitances of the interfaces. Couet et al. presented the coupled current charge compensation model in 2015 and successfully fitted it to experimental data [15]. The model does incorporate the transport of electrons but is based on several assumptions which are different from the presented study, among others a constant potential drop at the metal/film interface, no potential drop at the film/solution interface and no BTBT. In the same year, 2015, Momeni et al. introduced the mass-charge balance model. The influence of electrons is included by the Fermi-Levels and, among others, the driving force for the electrochemical reactions differs from the presented study [16].

The shown dependency of the electric field on the external potential in the previous study was derived from a model without considering the influence of the interactions of electrons and holes as well as the implications of energy levels on their transport and tunneling behavior. To analyze the effect of an inclusion of electrons and holes to modeling of oxide film growth and to test Macdonald's et al. hypothesis that the separation of electrons and holes shows a significant impact on the electric filed, we extend the model in this study by:
- Thermal generation of electrons and holes
- Recombination of electrons and holes
- BTBT of electrons from the valence band to the conduction band to generate electrons in the conduction band and holes in the valence band.
- Transport of electrons in the conduction band and holes in the valence band.

To test the impact of BTBT, we compare four different simulations, which are identical, with the exception of the behavior of electrons and holes.
The oxides only differ in the presence of charge carriers, their concentration (defined by the band gap) and the possibility of BTBT. The additionally parameters describing the oxide growth are identical; in this way it is possible to isolate the impact of electrons and holes, and BTBT on the oxide film growth.
We show that the case of possible BTBT can only be observed when the oxide film has very specific properties (thick oxide film with narrow band gap). In such cases, it does influence the electric field inside the oxide film and indeed has a buffering effect on the electric field. We also show that this effect does neither lead to a potential independent electric field nor to a uniform (space independent) electric field.

## 2. MODEL DESCRIPTION

A detailed description of the interfacial reactions, the transport of crystal defects and the calculation of the electric field by the Poisson equation and its boundary conditions is given in our earlier publication [13]. For the given case of additional charge carriers – electrons in the conduction band and holes in the valence band – the Poisson equation needs to be extended by these charge carriers:

$$\frac{\partial^2 \varphi}{\partial x^2} = \frac{\partial F_\mathrm{E}}{\partial x} = -\frac{F}{\varepsilon_\mathrm{r}\varepsilon_0}\sum z_i c_i \qquad (1)$$

whereas in this case $c_i$ includes not only crystal defects – namely oxygen and metal vacancies – but also electrons and holes. Accordingly, $z_i$ is the charge of the species and $\varepsilon_0$ is the permittivity of the vacuum 8.85×10$^{-12}$ As/V/m, $\varepsilon_\mathrm{r}$ describes the relative permittivity.

Further changes to the former model are described below.

## 2.1 Description of Electrons and Holes Transport and Generation

In addition to the transport of defects as described in our earlier publication, generation,

transport and boundary conditions of electrons and holes are an important concept for the modeling of the oxide film. It is assumed that the defects enter the oxide film as already charged defects. Thus, the number of electrons and holes is not directly affected by the number of defects as it can be expected in doped semiconductors. Of course, the number of defects affects the electric field which in turn has an impact on the transport equations and thus the concentration of electrons.

### 2.1.1 Transport equation

Electrons and holes can be transported by drift due to the electric field and by diffusion due to concentration gradients [17]. Both can be calculated equivalent to the transport of defects inside the film, leading to:

$$J_i = -D_i \frac{\partial c_i}{\partial x} - \frac{z_i F D_i}{RT} \frac{\partial \varphi}{\partial x} c_i \qquad (2)$$

with the concentration of either electrons or holes $c_i$, the diffusion coefficient of the the according charge carrier $D_i$, Faraday's constant $F$, the universal gas constant $R$ and the universal temperature $T$.

Additional to the transport of electrons and holes there will be thermal generation of electrons and holes and recombination of both, which leads to annihilation of the charge carriers. The rate of generation and recombination $r_G$ can be described by [18]:

$$r_G = \frac{c_e c_h - c_{i,0}^2}{\tau_R}. \qquad (3)$$

where the indexes e and h represent electrons and holes respectively and $c_{i,0}$ the concentration of intrinsic charge carriers. The conversation of mass is given for each species $i$ by:

$$\frac{\partial c_i}{\partial t} = -\Delta J_i + r_G. \qquad (4)$$

Even though, it is well known that many material parameters in oxide films can be position dependent, for example due to impurities or high doping, which might affect the transport of the charge carriers [19], for the purpose of this publication these effects are neglected.

### 2.1.2 Boundary conditions

*2.1.2.1 Metal/film interface*

The fermi level of electrons at the metal side and at the film site of the metal/film boundary must be in equilibrium. Assuming that the metal/film interface is a metallic junction, the fermi level at the metal side of the boundary can be calculated by the standard potential of the electrons $\mu_e^0$ in the metal and the external potential $\varphi_{\text{ext}}$ [17]:

$$E_F^m = \mu_e^0 - e\varphi_{\text{ext}} \tag{5}$$

Here $E_F^m$ is the fermi level of the electrons in the metal and $e$ the elementary charge. The concentration of electrons $c_e$ in the conduction band at the metal/film interface can be calculated by [18]:

$$c_e = N_c \exp\left(\frac{E_F - E_c}{k_B T}\right) \tag{6}$$

with the density of states $N_c$ in the conduction band, the fermi level $E_F$, the energy at the bottom edge of the conduction band $E_c$, the Boltzmann constant $k_B$ and the universal temperature $T$. Considering the equilibrium between fermi levels at the metal film interface $E_F^m = E_F^{m/f}$ and

$$E_c = E_c^0 - e\varphi \tag{7}$$

whereas $E_c^0$ is a fixed value, independent of electrical state and equal to the conduction band energy relative to vacuum. One can calculate the electrons' concentration at the metal film boundary as:

$$c_e = N_c \exp\left(\frac{\mu_e^0 - E_c^0 + (\varphi - \varphi_{\text{ext}})e}{k_B T}\right), \tag{8}$$

whereas $\varphi$ is the potential at the film side of the metal/film interface, or more precisely at the defect layer/film interface (compare [13] for a more detailed description of the potential distribution at the metal/film interface according to the R-PDM). It is debatable that electrons and holes can enter the defect layer, for numerical simplification the defect layer remains charge free in this study. The calculation of holes $c_h$ in the valence band at the metal/film interface can be done equivalent and leads to:

$$c_h = N_v \exp\left(\frac{E_v^0 - \mu_e^0}{k_B T}\right) \tag{9}$$

with the energy at the top edge of the valence band relative to vacuum $E_v^0$ and the density of states of holes $N_v$.

*2.1.2.2 Film/solution interface*

It is assumed that no electrochemical reaction is occurring at the film/solution interface. Thus, a no flux boundary condition applies at the film/solution interface for both, the electrons and the holes. The transport by diffusion needs to be balanced by the transport by migration which leads to:

$$-D_i \frac{dc_i}{dx} = z_i \mu_i c_i \frac{d\varphi}{dx} \tag{10}$$

### 2.1.3 Band-to-band Tunneling

Band-to-band tunneling (BTBT) can appear if the band bending of the conduction and the valence band is high enough to enable tunneling of electrons from one band to another without changing its energy. The mechanism of BTBT in contrast to thermal generation of electrons and holes is schematically shown in Figure 1. Without an electric field the energy of the valence band ($E_v$) and the conduction band ($E_c$) are independent of position in the oxide. With an electric field the band energy changes in space and $E_v$ and $E_c$ can have the same energy at different positions in the oxide film. This enables the tunneling of an electron from the valence band to the conduction band without a change in energy.

The rate of band-to-band tunneling $r_{BTB}$ can be described by [20,21]:

$$r_{BTB} = A \left(\frac{F_E}{F_0}\right)^P \exp\left(-\frac{B}{F_E}\right) \tag{11}$$

where $F_0 = 1\ \text{Vcm}^{-1}$, $P = 2$ for direct BTBT and $A$ and $B$ are the material specific Kane parameters and depend, among others, on the band gap, whereas a small band gap leads to a higher tunnel rate. It is important to note that in an infinite thick oxide film the smallest band bending will lead to BTBT, but in real oxide films with finite length a minimum electric field is needed to enable BTBT. The minimum electric field strength $F_E^{\min}$ can be calculated taking into account the change of band energy $E_i$:

$$E_i = E_i^0 - e\varphi \tag{12}$$

where the index $i$ represents either $c$ for the conduction band or $v$ in case of the valence band, $e$ is the elementary charge and $E_i^0$ is the band energy without an external potential. Thus, the change of band energy with space $dE_i/dx$ can be calculated by:

$$\frac{dE}{dx} = -e\frac{d\varphi}{dx} = -eF_E \tag{13}$$

Since the band bending (change of band energy) over the length of the oxide film must be at least equal to the energy of the band gap leading to $dE/dx = E_g/L$, the minimum field strength can be calculated as:

$$F_E^{\min} = \frac{E_g}{Le}. \tag{14}$$

For $F_E^{\min}$ electron tunneling from one edge to the other edge of the oxide film is possible, the higher the field strength compared to $F_E^{\min}$ the wider the interval in which BTBT is enabled. A calculation of the $F_E^{\min}$ depending on the band gap and the oxide film thickness shows that BTBT is only possible for thick oxide films, because also small band bending can be enough to bring two points of the valence band and the conduction band to the same energy, or oxides with very narrow band gap (Figure 2). The calculation of the tunnel distance $d_T$ can be done equivalent and yields:

$$d_T = \frac{E_g}{F_E e} \tag{15}$$

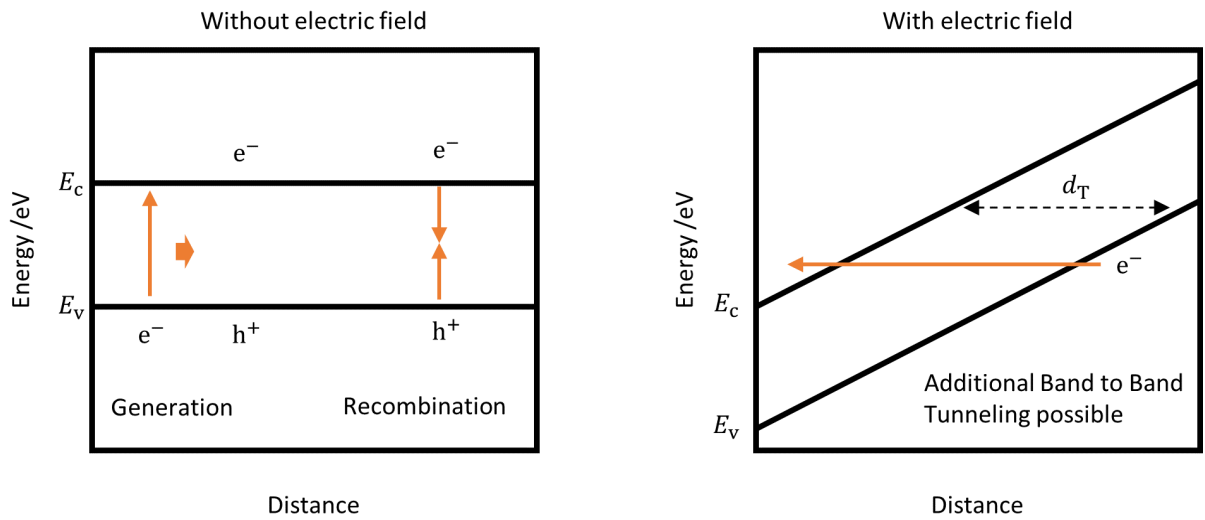

Figure 1: Schematic diagram of possible electrons and holes generation. $E_v$: energy of the valance band at the top edge of the band, $E_c$: energy of the conduction band at the lower edge. $d_T$: tunnel distance.

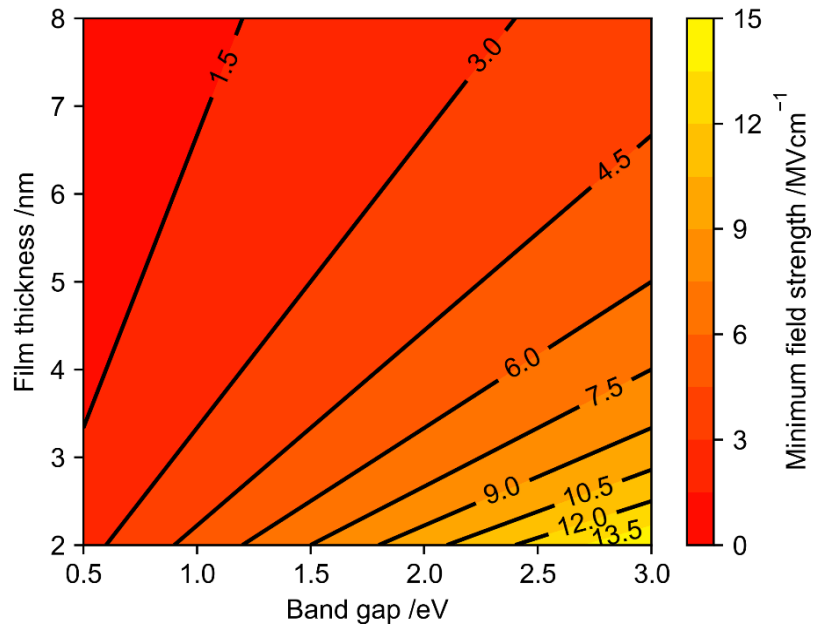

Figure 2: Minimum field strength to enable band-to-band tunneling from one edge of the film to another depending on the film thickness and the band gap.

## 2.2 Case study Parameters

Additionally, to the parameters given in the previous publication on oxide film growth publication, the transport and generation/recombination of electrons must be described by several variables/constant, which are given in Table 1.

Table 1: Model parameters

| Symbol | Model parameter | Parameter value | Parameter unit |
| --- | --- | --- | --- |
| $e$ | Elementary charge | $1.602 \times 10^{-19}$ | C |
| $N_A$ | Avogadro constant | $6.022 \times 10^{23}$ | 1/mol |
| $N_c$ | Density of states in the conduction band | $1 \times 10^{20}/N_A$ | mol/cm$^3$ |
| $N_v$ | Density of states valence band | $N_c$ | mol/cm$^3$ |
| $k_B$ | Boltzmann constant | $1.38 \times 10^{23}$ | J/K |
| $\mu_e$ | EC potential electron with respect to vacuum | 2.0 | eV |
| $E_F^m$ | Fermi level metal | $\mu_e^0 - e\varphi_{ext}$ | eV |
| $E_c^0$ | Energy band edge conduction band without external potential | 2.25 / 3.0 | eV |
| $E_v^0$ | Energy band edge valence band without external potential | 1.75 / 1.0 | eV |
| $\tau_R$ | Recombination constant | $3 \times 10^5/N_A$ | mols/cm$^3$ |
| $D_e$ | Diffusion coefficient electron | $3.28 \times 10^{-4}$ | m$^2$/s |
| $D_h$ | Diffusion coefficient holes | $3.28 \times 10^{-4}$ | m$^2$/s |
| $A$ | First Kane parameter for BTBT | $1.4 \times 10^{20}/N_A$ | mol/m$^3$/s |
| $B$ | Second Kane parameter for BTBT | 5 | MV/cm |
| $z_e$ | Charge number electron | $-1$ | |
| $z_h$ | Charge number hole | 1 | |
| $r_G$ | Generation rate | $1 \times 10^{-10}$ | mol/m$^3$/s |

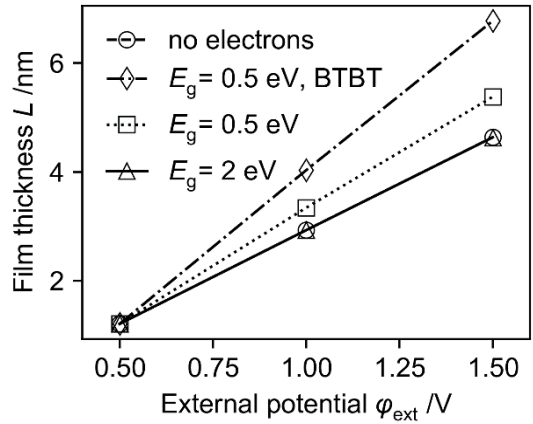

Figure 3: Oxide film thickness at steady state depending on the external potential calculated for four different cases: without inclusion of electrons, wide band gap (2 eV) and no BTBT, narrow band gap (0.5 eV) and no BTBT and narrow band gap (0.5 eV) and BTBT.

## 3. INFLUENCE OF ELECTRONS AND HOLES ON STEADY STATE OXIDE FILM PROPERTIES

The influence of electrons and holes on the behavior of oxide films strongly depends on the number of electrons in the conduction band and the number of holes in the valence band. The following calculations are separated in four different cases:
1. An oxide film with a band gap of 2 eV, which is in the range of iron oxide [22], and thus does not tend to BTBT for field strengths of around 3 MV/cm and film thicknesses of around 5 nm.
2. Artificial oxide films are calculated with narrower band gap of 0.5 eV with high BTBT (defined by a high value of Kanes parameter *A*)
3. As before, but without BTBT.
4. Base case without any inclusion of electrons and holes at all.

The calculations are performed using Equations (1-10) from the previous manuscript [13] in combination with all equations listed above.

Assuming a wide band gap and thus a low concentration of electrons (as can be seen from Equation 5) the influence of electrons and holes on the film thickness of oxide film is neglectable (Figure 3). Due to a narrow band gap, or more precisely due to a smaller difference $E_\mathrm{F} - E_\mathrm{c}$ and $E_\mathrm{v} - E_\mathrm{F}$, the number of electrons and holes increases and their effect on the steady state film thickness becomes more pronounced. With additional BTBT the effect is even more visible. Due to increased BTBT with increasing field strength, the effect of a thicker steady state film thickness is more pronounced at higher external potentials. It should be noted that the effect of BTBT on the steady state film thickness (and on the additional oxide film behavior) strongly depends on the Kane parameters (and thus the rate of BTBT). In this case a high value for $A$ is chosen to clarify this effect. As discussed further in the manuscript, BTBT affects strongly the electric field distribution, by influencing the accumulation of charge at the metal/film and film/solution interfaces. This is due to the fact that the BTBT enhance the separation of charge in the film: it brings the electrons close to the metal/film interface and the holes close to the film/solution interface. So, while in the film the electric field is buffered, at its extreme it is enhanced.

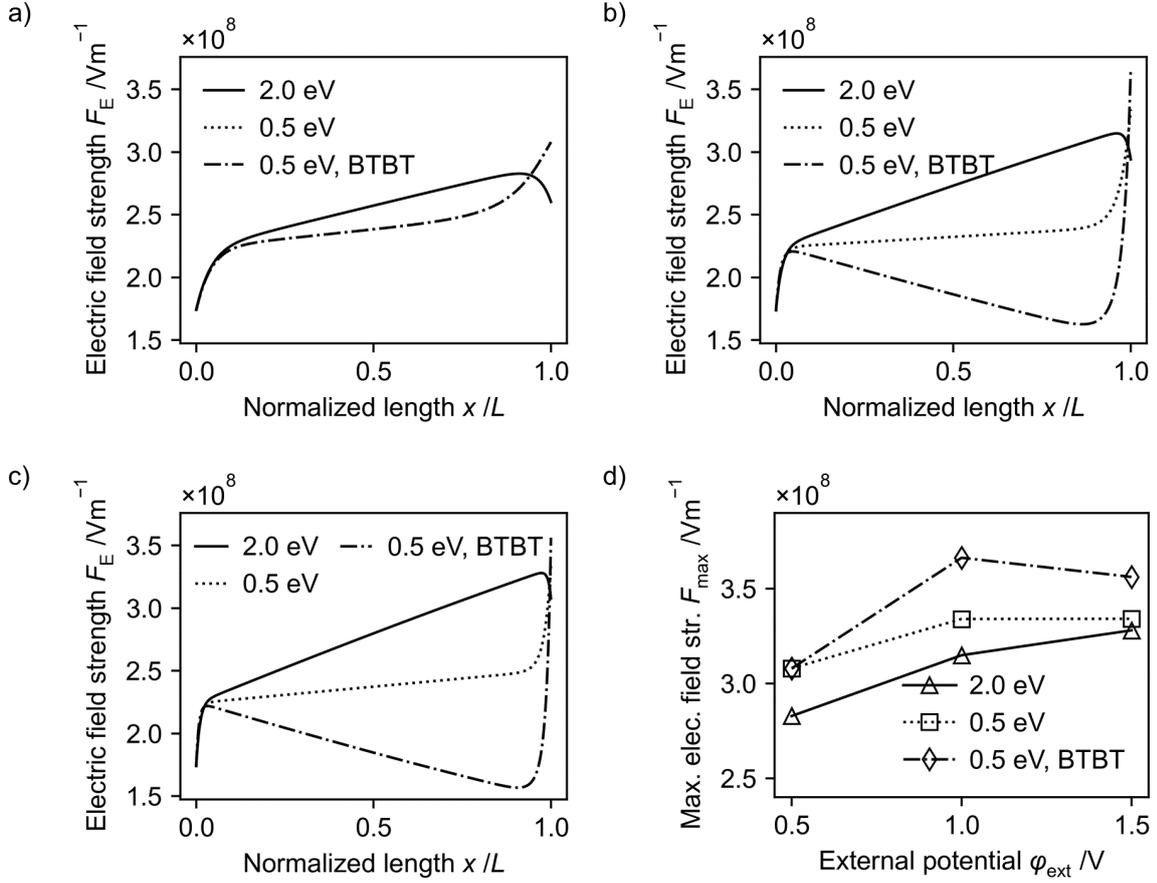

Figure 4: Electric field strength for three different cases – wide band gap (2 eV), narrow band gap (0.5 eV) without and with BTBT - depending on the external potential. a) Electric field strength calculated for an external potential of 0.5 V; b) 1.0 V; c) 1.5 V; d) Maximum electric field strength depending on the external potential.

The electric field strength is a matter of debate in case of oxide film modeling [10,12,23]. It has been assumed the field strength will be buffered by BTBT and the separation of electrons and holes [2]. There is indeed a buffering effect visible due to the influence of electrons and band-to-band tunneling (Figure 4). At a low external potential, the electric field is not high enough to enable BTBT but the higher number of electrons, due to the narrow band gap, leads to a lowering of the electric field over a broad range of the oxide film. At the film solution interface, the electric field increases sharply due to the accumulation of holes (compare below). Both cases with narrow band gap are similar (because no BTBT occurs) and differ slightly from the case with broad band gap (which is nearly identically to the case without electrons (not shown)) (Figure 4a). With increasing external potential from 1.0 V (Figure 4b) to 1.5 V (Figure 4c) the effect becomes more pronounced and is even reinforced by the BTBT.

A comparison of the maximum electric field strength also shows the buffering effect of electrons and holes on the electric field (Figure 4d). Nevertheless, it is important to note that the effect neither leads to a potential independent electric field nor to a uniform space independent electric field.

The conentration of electrons and holes is affected by the electric field and by the possibility of BTBT (Figure 5). Due to the electric field the electrons are driven towards the metal/film interface and the holes are driven towards the film/solution interface.

The Fermilevel of the metal in contact with the oxide film determines the electrons concentration at the boundary. If no tunnel current is possible, the electrons concentration decreases with distance to the metal/film interface (Figure 5a). The electric field drives the holes in the opposite direction. At the film/solution boundary a accumulation of holes is visible due to no flux from the oxide to the solution and the assumption of no electrochemical reaction (Figure 5b). Due to the assumption of no electrochemical reactions at the film/solution boundary (and thus the choice of boundary conditions) the total amount of holes in the oxide is drastically higher compared to the amount of electrons. The holes are only balanced by the recombination with the electrons (which is very low at the film/solution interface due to the transport by the electric field). The electrons, in turn, are also balanced by the fermi energy at the metal/film interface. Figure 5a and 5b show the case for a narrow band gap (0.5 eV), in case of the broader band gap (2 eV) the concentration profile is similar with lower total concentration of charge carriers (not shown).

Enabling the possibility of BTBT increases the amount of electrons and leads to a new concentration profile (Figure 5c). The concentration at the metal/film interface is still dictated by the dirichlet boundary condition but a high flux of electrons from the valence band near the film/solution interface to the conduction band near the metal/film interface (compare Figure 1) leads to an accumulation of electrons. It is clearly visible at which point the electrons start to enter the conduction band by a steep increase of concentration.

In case of the holes the generation of electrons in the conduction band near the metal/film interface causes a generation of holes in the valence band near the filmn/solution interface. This leads – in combination with the zero flux boundary condition – to an even increased concentration of holes. This accumulation can only be buffered by the very low concentration of electrons due to the recombination mechanism. An oxidation reaction at this interface would be strongly enhanced by the flux of holes from the oxide to the solution and would lead to a lower holes concentration and buffer the effect of this charge accumulation and thus buffer the electric field at the film/solution interface. The strenght of this effect will strongly depend on the reaction rate of the electrochemical reaction consuming the holes from the oxide film. Assuming a reaction rate comparable to the reaction rates of vacancies production ($10^{-8}$ mol/m$^{-2}$s$^{-1}$) the effect is negligible. Nevertheless, assuming an oxidation of the film itself by a reaction with the holes would eventually (under high enough potentials and thus high h$^+$ concentrations) lead to transpassive breakdown of the film. Due to the enhanced rate of holes generation BTBT will increase the rate of electrochemical reactions at the film solution interface.

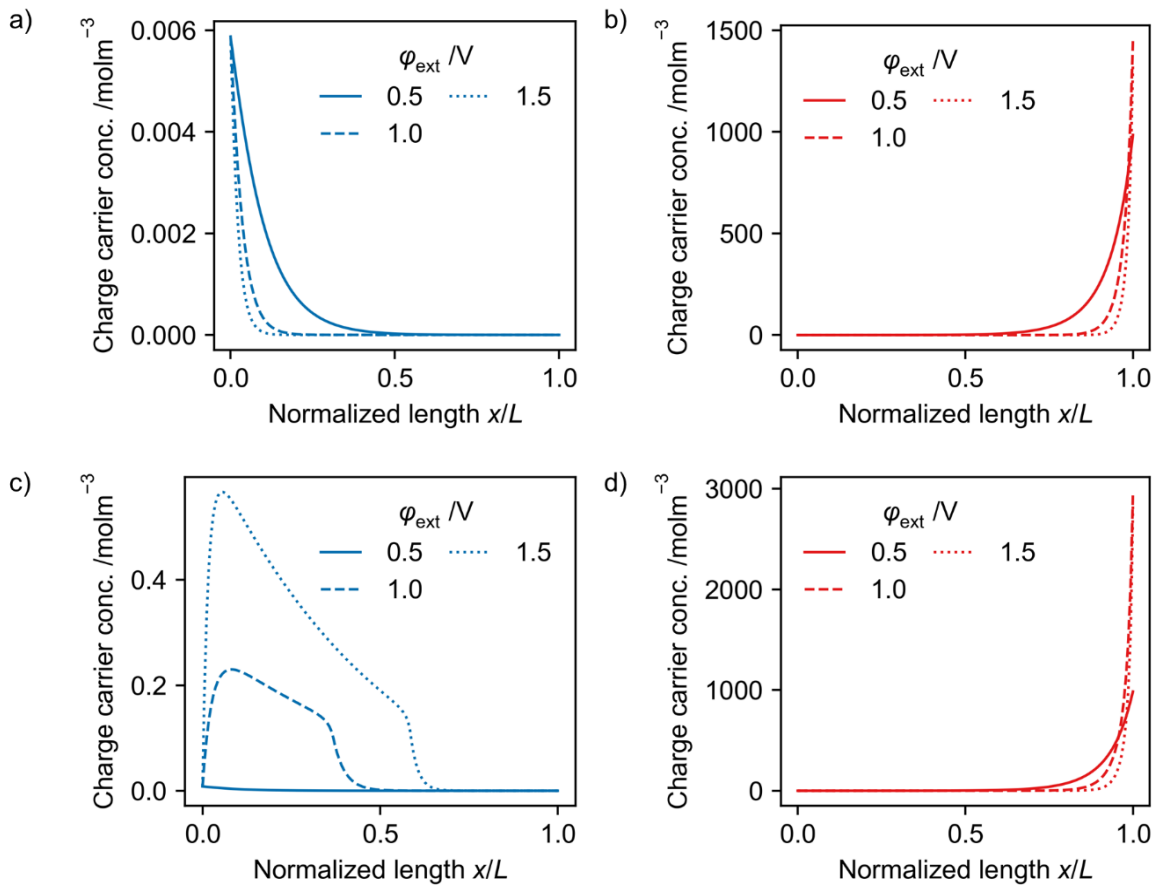

Figure 5: Concentration of charge carriers over the normalized film length depending on the external potential. a) concentration of electrons (blue) and b) holes (red) without BTBT. c) electrons and d) holes concentration for enabled BTBT; $E_G = 0.5$ eV in all cases.

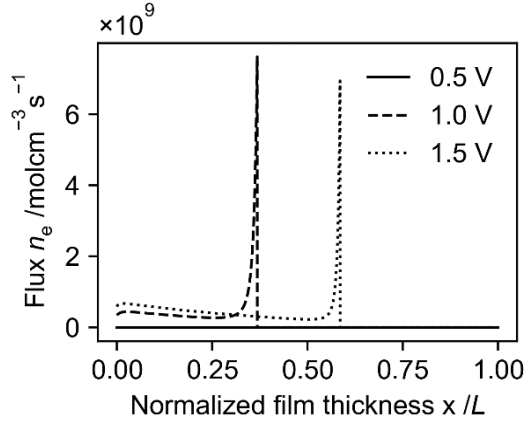

Figure 6: Electron flux due to band-to-band tunneling depending on the external potential. The flux describes the flux of electrons to the conduction band from the valence band near the film/solution interface.

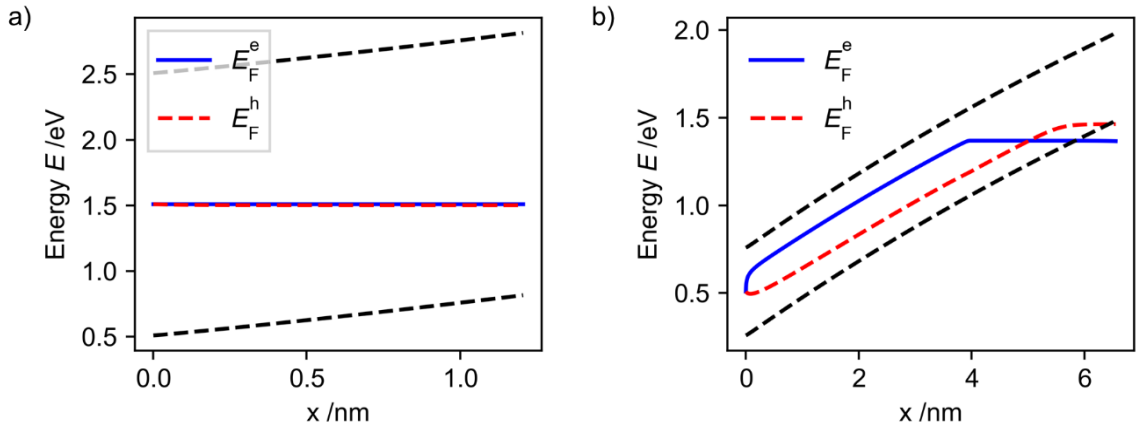

Figure 7: Energy levels and Quasi Fermi Levels calculated for an external potential of 1.0 V; a) Band gap of 2 eV – no band-to-band tunneling; b) Band gap of 0.5 eV – band-to-band tunneling possible.

The tunnel flux of electrons to the conduction band at position $x$ is determined by the electric field at position $x + d_\text{T}$, where $d_\text{T}$ is the tunnel distance (compare Figure 1). The electron flux due to band-to-band tunneling is very homogenous over a broad range at the beginning due to the homogenous electric field strength (Figure 6). At higher external potentials the interval of tunnel flux is enlarged due to a thicker film. In addition, the sharp increase of the electric field at the film/solution interface causes a sharp increase of the flux of electrons when $x + d_\text{T}$ is near the film/solution interface, which drops to zero when $x + d_\text{T} > L$.

The band energy and quasi fermi level calculated by the electrons via:

$$E_F^e = E_c + \ln\left(\frac{c_e}{N_c}\right) k_B T \tag{16}$$

and by the holes calculated by:

$$E_F^h = E_v - \ln\left(\frac{c_h}{N_v}\right) k_B T \tag{17}$$

The energy of the band edges of the conduction band can be calculated by Equation 6 and the energy of the band edge of the valence band by $E_v = E_v^0 - \varphi e$.

The band energy and Quasi Femi Level for an external Potential of 1.0 V and a band gap of 2 eV (and thus without BTBT) and with a narrow band gap of 0.5 eV and with BTBT is given in Figure 7a and Figure 7b, respectively. Due to the decreasing potential the energy of the conduction band edge (upper dashed black line) and the energy of the valence band edge (lower dashed black line) increase nearly linearly. Without BTBT the concentration of holes and electrons show a logarithmic distribution. Thus, the calculated quasi fermi level is constant (Figure 7a). If BTBT is enabled, the concentration of electrons and holes show a different distribution. Thus, the quasi-fermi levels differ from each other and from the case without BTBT (Figure 7b).

## 4. CONCLUSION

The effect of electrons and holes on the modeling of steady state oxide films based on a refined PDM (R-PDM) has been calculated. It has been shown that the influence of these charge carriers can be neglected in case of broad band gaps and low concentrations of electrons and holes if no electrochemical reaction is expected at the film solution interface.

In case of narrow band gaps and higher charge carrier concentrations the effect of electrons and holes must be considered to the calculation of oxide films. A high electrons' and holes' concentration can have a buffering effect on the electric field and influences the potential at the film/solution interface, which could - among others - influence electrochemical reactions at this interface.

In contrast to previous assumptions the buffering effect does not lead to an independency of the electric field on the external potential. It can even lead to a lowering of the electric field with increasing potential over broad ranges of the oxide film but leads to a sharp increase of the electric field strength at the film/solution interface. This behavior strongly depends on the chosen parameters for the oxide film modeling.

By our theoretical considerations on the transport and generation of electrons and holes as well as BTBT it could be shown, that the PDM assumption of a uniform, potential independent electric field is not generally valid. For a more detailed description of oxide film growth a more complex model is necessary, considering the transport of charged species and the Poisson equation.

For the fitting of experimental EIS data to oxide film models it is important to add all mechanism that affect the oxide film behavior. Only on this way the fitted kinetic data will be a good reflection of reality and can be trusted for predictions on the oxide behavior. Especially, in case of electrochemical reactions at the film/solution interface, like film dissolution and breakdown, the inclusion of electrons and holes seems to be inevitable for a deeper understanding of the processes and a prediction of the corrosion behavior.

A first step for the modeling of oxide films could be an assessment of the expected electrons' concentrations by the potential energy of the electrons, the fermi level of the metal and the band energy of the conduction band. Furthermore, the possibility of BTBT can be estimated by the steady state film thickness and the band gap. If the band gap of the oxide is large, the impact of electrons and holes on the film growth will be negligible, due to their low concentration. In any case, even under such circumstances, as shown in part one of this publication [13], the model should include a calculation of the electric field by the Poisson equation because it can neither assumed as constant regarding the position inside the oxide, nor regarding the external potential, and it has a strong effect on the oxide film properties.

## 5. ACKNOWLEDGEMENT

The authors would like to thank Georg Pesch for his patience in going through equation after equation together for a few afternoons and searching for sign errors.

## 6. REFERENCES

[1] C.Y. Chao, L.F. Lin, D.D. Macdonald, A Point Defect Model for Anodic Passive Films, *J. Electrochem. Soc.* 128 (1981) 1187.

[2] D.D. MacDonald, The history of the Point Defect Model for the passive state: A brief review of film growth aspects, *Electrochim. Acta.* 56 (2011) 1761–1772.

[3] S. Sharifi-Asl, M.L. Taylor, Z. Lu, G.R. Engelhardt, B. Kursten, D.D. Macdonald, Modeling of the electrochemical impedance spectroscopic behavior of passive iron using a genetic algorithm approach, *Electrochim. Acta.* 102 (2013) 161–173.

[4] D. Macdonald, S. Sharifi-Asl, G. Engelhardt, Review of the extraction of electrochemical kinetic data from electrochemical impedance data using genetic algorithm optimization, *Bulg. Chem. Commun.* 49 (2017) 53–64.

[5] I. Nicic, D.D. Macdonald, The passivity of Type 316L stainless steel in borate buffer solution, *J. Nucl. Mater.* 379 (2008) 54–58.

[6] D.D. Macdonald, A. Sun, N. Priyantha, P. Jayaweera, An electrochemical impedance study of Alloy-22 in NaCl brine at elevated temperature: II. Reaction mechanism analysis, *J. Electroanal. Chem.* 572 (2004) 421–431.

[7] A. Fattah-alhosseini, F. Soltani, F. Shirsalimi, B. Ezadi, N. Attarzadeh, The semiconducting properties of passive films formed on AISI 316 L and AISI 321 stainless steels: A test of the point defect model (PDM), *Corros. Sci.* 53 (2011) 3186–3192.

[8] C.-O.A. Olsson, D. Hamm, D. Landolt, Evaluation of Passive Film Growth Models with the Electrochemical Quartz Crystal Microbalance on PVD Deposited Cr, *J. Electrochem. Soc.* 147 (2002) 4093.

[9] I. Bösing, G. Marquardt, J. Thöming, Effect of Heat Treatment of Martensitic Stainless Steel on Passive Layer Growth Kinetics Studied by Electrochemical Impedance Spectroscopy in Conjunction with the Point Defect Model, *Corros. Mater. Degrad.* (2020) 1–15.

[10] M. Vankeerberghen, 1D steady-state finite-element modelling of a bi-carrier one-layer oxide film, *Corros. Sci.* 48 (2006) 3609–3628.

[11] A. Seyeux, V. Maurice, P. Marcus, Oxide Film Growth Kinetics on Metals and Alloys: I. Physical Model, *J. Electrochem. Soc.* 160 (2013) C189–C196.

[12] G.R. Engelhardt, B. Kursten, D.D. Macdonald, On the nature of the electric field within the barrier layer of a passive film, *Electrochim. Acta*. 313 (2019) 367–377.

[13] I. Bösing, F. La Mantia, J. Thöming, Modeling Electrochemical Oxide Film Growth – A PDM Refinement, *Electrochem. Acta*. ?? (2021) ??

[14] C. Bataillon, F. Bouchon, C. Chainais-Hillairet, C. Desgranges, E. Hoarau, F. Martin, S. Perrin, M. Tupin, J. Talandier, Corrosion modelling of iron based alloy in nuclear waste repository, *Electrochim. Acta*. 55 (2010) 4451–4467.

[15] A. Couet, A.T. Motta, A. Ambard, The coupled current charge compensation model for zirconium alloy fuel cladding oxidation: I. Parabolic oxidation of zirconium alloys, *Corros. Sci.* 100 (2015) 73–84.

[16] M. Momeni, J.C. Wren, A mechanistic model for oxide growth and dissolution during corrosion of Cr- containing alloys †, *Faraday Discuss.* 00 (2015) 1–23.

[17] J. Newman, K.E. Thomas-Alyea, J. Newman, K.E. Thomas-Alyea, Electrochemical Systems, 3rd ed., John Wiley & Sons, (2012).

[18] F. Thuselt, F. Thuselt, Physik der Halbleiterbauelemente, (2011).

[19] A.H. Marshak, Modeling semiconductor devices with position-dependent material parameters, *IEEE Trans. Electron Devices*. 36 (1989) 1764–1772.

[20] E.O. Kane, Zener tunneling in semiconductors, *J. Phys. Chem. Solids*. 12 (1960) 181–188.

[21] K.H. Kao, A.S. Verhulst, W.G. Vandenberghe, B. Soree, G. Groeseneken, K. De Meyer, Direct and indirect band-to-band tunneling in germanium-based TFETs, *IEEE Trans. Electron Devices*. 59 (2012) 292–301.

[22] J.K. Leland, A.J. Bard, Photochemistry of colloidal semiconducting iron oxide polymorphs, *J. Phys. Chem.* 91 (1987) 5076–5083.

[23] B. Krishnamurthy, R.E. White, H.J. Ploehn, Electric field strength effects on time-dependent passi v ation of metal surfaces, 47 (2002).

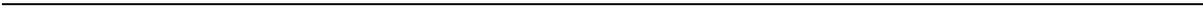